\documentclass[submission,copyright,creativecommons]{eptcs}
\providecommand{\event}{TFPIE 2015} % Name of the event you are submitting to
\usepackage{breakurl}             % Not needed if you use pdflatex only.
\usepackage{listings}

\usepackage[usenames,dvipsnames]{color}

\title{Teaching Functional Patterns through Robotic Applications}
\author{J. Boender, E. Currie, M. Loomes, G. Primiero, F. Raimondi
\institute{School of Science and Technology\\ 
Middlesex University, London}
\email{\{j.boender,e.currie,m.loomes,g.primiero,f.raimondi\}@mdx.ac.uk}
}

\def\authorrunning{Boender, Currie, Loomes, Primiero \& Raimondi}
\def\titlerunning{Teaching Functional Patterns through Robotic Applications}
\begin{document}
\maketitle

\begin{abstract} 
  We present our approach to teaching functional programming to First Year Computer Science students at Middlesex University through projects in robotics. A holistic approach is taken to the curriculum, emphasising the connections between different subject areas. A key part of the students' learning is through practical projects that draw upon and integrate the taught material. To support these, we developed the Middlesex Robotic plaTfOrm (MIRTO), an open-source platform built using Raspberry Pi, Arduino, HUB-ee wheels and running Racket (a LISP dialect). In this paper we present the motivations for our choices and explain how a number of concepts of functional programming may be employed when programming robotic applications. We present some students' work with robotics projects: we consider the use of robotics projects to have been a success, both for their value in reinforcing students' understanding of programming concepts and for their value in motivating the students.  \end{abstract}

\section{Introduction}\label{Section:intro}

%  Motivate students with practical and "physical" manifestation of software: let's use a robot.
 % This is in the context of a new programme etc., FP because:
% Close to maths formalism
% All students start from the same point (zero knowledge)

%  In another paper we showed that we can use Racket.
 % Here we present functional patterns for robotic applications

This paper discusses how the language Racket has been used in the first year of the Computer Science programme at Middlesex University, with a focus on the use of physical devices and robotics to teach aspects of functional and imperative programming and to reinforce other areas of the curriculum. The background lies in the development of a new BSc CS programme, which has now reached the end of its second year, so that the first year has seen now two cohorts of students. The first year of the programme takes a holistic approach to providing a solid grounding in computer science; there are no modules, but rather a number of interwoven themes, namely programming, physical computing, formal underpinnings, design and project work. The approach involves exposing students to key concepts in each of these areas. Taking propositional logic as an example, there is a theoretical treatment in the formal sessions, practical implementation of logic formulae with gates in the physical computing sessions, implementation as boolean functions in programming labs, modelling the language in design sessions and application of the above in project work.

One of the key decisions in the design of this programme was the choice of the programming language. Racket was chosen because it could be used as the `glue' to hold together the other parts of the programme. Many of the concepts covered on the course can be implemented in Racket and this language proved ideal for interfacing with microcontrollers and robots in the integrative project work. It was decided at an early stage that we would try to motivate students and draw together the various topics by having them engage in projects that involved practical and `physical' manifestations of software. The academic year was divided into three blocks and each of these had an associated project. The first block project involved the use of an Arduino micro-controller controlled using Racket. For the first cohort, the project was the design of a 3-way traffic light system for roadworks; the second cohort used an LED matrix to implement a noughts and crosses game. The second block focused on data structures, and the associated projects involved the design of a `dungeon' game. For the third block, the students did projects based on a robot developed in-house and this is the main topic of the paper, as described below.

These projects enabled students to apply and integrate a number of topics from other areas of the curriculum. For example, they used finite state machines to describe the required mutual behaviour of the robot wheel motors. As discussed in Section~\ref{Section:patterns}, students used propositional logic functions implemented in Racket for tasks such as verifying that a proposed speed was within a robot's designated range. Principles from the design and formal underpinnings sessions were applied in creating new applications for the robots; for example, open- versus closed-loop feedback systems. Finally, the projects were carried out in groups, which developed the students' associated transferable skills.

The rest of the paper is organised as follows. We first provide an overview of our robotic platform in Section~\ref{Section:RacketMirto}. In Section~\ref{Section:patterns} we describe the patterns that we have observed and taught in the programme, and in Section~\ref{Section:examples} we present examples of students' projects. We discuss related literature in Section~\ref{Section:lit}.

\section{Overview of Racket and MIRTO}\label{Section:RacketMirto}

As with all choices of programming language, our choice of Racket was a compromise. Perhaps the major factor in our decision was that Racket could be used as a unifying notation with which to explore all of the first year material; because it is also an imperative language, we could also use it to cover the concepts of state and iteration with loops that the students would meet in their second year work with Java; and because of its functional flavour, we could use it to highlight some of the logic notions recurring in all other contexts. 

A convenient feature of Racket is that all the imperative `functions' (procedures) in the language have names that end with an exclamation mark (!). Thus students can be aware when they are programming imperatively, and if they want to use a purely functional style, they can do so by not using these functions. The ability to use `functions' that return void and do their tasks by side effects adds the flexibility needed for many of the robot-controlling functions used in the course, while of course also helping students to learn about side effects. Therefore, while not for the functional programming purist, the flexibility and range of Racket made it an ideal first language for our CS programme.

Some features of functional programming are not so easy in Racket. For example, the use of infinite data structures is difficult because the language uses eager evaluation. However, the practical nature of the first year meant that these more esoteric aspects of functional programming were not as important as the flexibility of the language for a range of practical projects. To this aim, the Computer Science Department at Middlesex University, in collaboration with the Design, Engineering and Mathematics Department, have developed MIRTO (Middlesex Robotic plaTfOrm), a flexible open-source platform; its current design and all the source code are available on-line \cite{mirtogit}. Mirto is composed of two units:

\begin{enumerate}
\item The base platform provides two HUB-ee wheels \cite{hubwheels}, which include motors and encoders (to measure actual rotation) built in, a rechargeable battery pack, front and rear castors, two bump sensors and an array of six infra-red sensors (mounted under the base), and an Arduino microcontroller board with shield to interface to all of these.

\item The top layer consists of a Raspberry Pi, running a bespoke Linux image extending the standard Raspbian image, with Racket 6.1 installed and connected to the Arduino by the serial port available on its interface connection. 
\end{enumerate}
The control and monitoring of the micro-controllers is obtained through running the Arduino Service Interface Protocol (ASIP), a protocol similar, in spirit, to the Firmata protocol \cite{Steiner2009} in that it enables a computer to discover, configure, read and write a microcontroller's general purpose IO pins. However, ASIP has a smaller footprint than Firmata (using around 20\% less RAM) and it supports high level abstractions that can be easily attached to hundreds of different services for accessing sensors or controlling actuators. These abstractions can decouple references to specific hardware, thus enabling different microcontrollers to be used without software modification. For an overview of the ASIP protocol see \cite{sescps15}. The Racket ASIP client library is available at \cite{asipt_racket} together with implementations for Input-Output, distance, motor with encoders, Infra-red sensors for line following, and NeoPixels services. The following is an example of Racket code to set pins $11, 12$ and $13$ of the Arduino board to HIGH: 

\begin{lstlisting} 
(map (lambda (x) (digital-write x HIGH))(list 11 12 13))
\end{lstlisting}
The code above makes use of the higher-order function \texttt{map}, applied to a $\lambda$-function which applies the ASIP library function digital-write to the list of numbers $11,12$ and $13$. As already shown in this short example, Racket provides an opportunity to teach functional programming languages in physical computing sessions.

An additional advantage of the setup with the Arduino and the Raspberry Pi is that it can be used to teach several other important concepts. For example, we use the Arduino to teach some elementary assembly programming (Atmel Studio~\cite{atmel} is an excellent simulator and IDE, currently in use at Middlesex). Additionally, we can teach some rudimentary Linux skills as well, such as command line operations.

%NOTE challenge winners etc.

%Show videos at presentation.

\section{Functional patterns for robots}\label{Section:patterns}

Our work with MIRTO robots induces the use of functional programming patterns by students. The philosophy of the course is for students to explore ideas and learn abstract concepts by a process of practical guided discovery, with the role of the tutors as facilitators. Students' understanding is deepened in a `spiral curriculum' approach by applying previously covered ideas in their project work. The approach is supported by the interactive nature of Racket, which enables students to try things quickly to explore why they get particular results. Students' understanding of concepts and their implementation is deepened by returning to the concept in a new context, either in a different subject area or by applying it in their project work. The project work is undertaken in groups, and the groups present their work to each other, which promotes peer learning.

\subsection{Random application of functions from a list and side effects}\label{section:random}

A first example is the exploration of the concept of side effects and the difference between symbols and their evaluation through a number of small exercises. While initially such exercises can be rather abstract and the understanding gained can be shallow and transient, our students returned to these concepts when they were asked to use the Racket library for ASIP to make a MIRTO robot explore an unknown area, as follows:

\begin{itemize}
	\item The robot should start by moving forward.
	\item When the robot hits an obstacle, it should stop immediately and move backwards for 0.5 seconds
	\item At this point, the robot should perform a left or a right rotation (randomly), and then restart and move forward until it hits the next obstacle.
\end{itemize}
As an additional feature, the time for the rotation was also to be random, say between 0.3 and 1.5 seconds, although we will ignore this aspect here.
To provide a solution for this exercise, a group of students wrote two functions, one to rotate left and one to rotate right, something similar to the following:

\begin{lstlisting} 
(define moveLeft
 (lambda ()
 ;; code here to move left, using the 
 ;; racket-asip library
 (printf "The robot moves left \n")
 )
 )
 
(define moveRight
 (lambda ()
 ;; code here to move right, using the 
 ;; racket-asip library
 (printf "The robot moves right \n")
 )
 )
 
(list-ref (list (moveLeft) (moveRight)) (random 2))
\end{lstlisting}

We abstract here from the details of the Racket-Asip library, as the key point here is the last line: the students defined a list of two functions with
{\tt (list (moveLeft) (moveRight))} and then used \texttt{list-ref} to get one element from this list at a position which is randomly $0$ or $1$, depending on the result of \verb+(random 2)+. They independently came up with a neat solution, and were clearly thinking `in a functional style' when defining a list of functions. There is, however, a problem with the code above; running it causes the robot to move both left and right, as both functions moveLeft and moveRight are executed. This led to an interesting seminar discussion that helped to deepen students' understanding in several areas. The problem was that writing \verb+(list (moveLeft) (moveRight))+ produces a list that contains the \emph{result} of invoking \texttt{moveLeft} and \texttt{moveRight}; Racket's eager evaluation means that both arguments to the function list are evaluated before it is applied. The functions have the side effect of printing to the screen. The contents of the list are the void values returned by the two functions (because \texttt{printf} returns void), and as a result \texttt{list-ref} chooses a random value from a list of voids. The solution provided at the end of the discussion is to build a list of references to the functions \texttt{moveLeft} and \texttt{moveRight}, rather than applications of them, by removing the brackets around them: 

\begin{lstlisting}
(list-ref (list moveLeft moveRight) (random 2))
\end{lstlisting}
This code will sometimes return a reference to \texttt{moveLeft}, and sometimes a reference to \texttt{moveRight}. To execute this reference, we need to surround the \texttt{list-ref} command with another pair of brackets. 

\begin{lstlisting}
((list-ref (list moveLeft moveRight) (random 2)))
\end{lstlisting}

The point is that this idea was generated by the students' own desire to make their robot do something interesting. Without this motivation, it is unlikely that they would have explored the concepts in sufficient detail to produce their proposed solution and, in turn, stimulate further discussion about how to make it work, which deepened their understanding of side effects and the difference between a symbol and its evaluation.

\subsection{Using higher order functions}

There are a number of instances in the projects where students may apply the programming concepts they have learned. One concept that many students find challenging is higher order functions. There are a number of possible ways to deploy higher order functions in controlling a robot with Racket, which enable students to see the concept applied in practical situations. As an example, we will consider how the Racket client for ASIP can be used to process Arduino analog input pins. The relevant input message received from the Arduino is a string of the following form: 
$$
\verb+@I,a,3,{0:320,1:340,2:329}+
$$
This indicates that these are analog pins, $3$ of which are set, pin $0$ to $320$, pin $1$ to $340$ and pin $2$ to $329$. A vector \texttt{ANALOG-IO-PINS} is defined to hold the values of the pins:

\begin{lstlisting}
(define MAX_NUM_ANALOG_PINS 16) 
(define ANALOG-IO-PINS (make-vector MAX_NUM_ANALOG_PINS))
\end{lstlisting}

and the code to update the vector is as follows:

\begin{lstlisting}
(define (process-analog-values input))

 (define analogValues (string-split (substring input 
                           (+ (str-index-of input "{") 1)
                           (str-index-of input "}") ) ",") )

(map (lambda (x) (vector-set! ANALOG-IO-PINS
         (string->number (first (string-split x ":")))  ;; the pin
         (string->number (second (string-split x ":"))) ;; the value
         ) ) ;; end of lambda
   analogValues) ;; end of map
 (printf "The current value of analog pins is: ~a \n" ANALOG-IO-PINS)
  ) ;; end process-analog-values
\end{lstlisting}

First we obtain the substring of the input message between the braces (\texttt{str-index-of} is defined below) and split to obtain the list \texttt{analogValues} of the form $("0:320" "1:340"...)$. We then map a function to set an analog pin to a given value, over the list of pin/value pairs. \texttt{str-index-of} is a utility function to find the index of a character $x$ in a string \texttt{str}; $x$ needs to be a string although we only look for its first character.

\begin{lstlisting}
(define (str-index-of str x)
(define l (string->list str))
(for/or ([y l] [i (in-naturals)] #:when (equal? (string-ref x 0) y)) i))
\end{lstlisting}
Working with the above gave students further practice both with the imperative features of Racket and with higher order functions and string processing. Some students would develop a deep understanding of the code, while others might gain a superficial understanding sufficient to use the code. The main benefit was for students to get used to working with and taking advantage of code that they hadn't written and that tested their ability to learn and to understand; in other words, to get a feel for programming in the real world. 

As a further example, here is some code using map in a simple control loop to print the value of the robot's IR sensors every 3 seconds and to print when the bump sensors are pressed or released:

\begin{lstlisting}
(define previousTime (current-inexact-milliseconds))
(define currentTime 0)

;; How often should we print?
(define interval 3000)

;; The list of IR sensors (numbered 0,1,2 and used in map below)
(define irSensors (list 0 1 2))

(define previousLeft #f)
(define previousRight #f)

(define (controlLoop)
  
  (set! currentTime (current-inexact-milliseconds))

  ;; Print IR values
  (cond ( (> (- currentTime previousTime) interval)
     ;; We use map to print the value of each sensor
     (map (lambda (i) (printf "IR sensor ~a -> ~a; " i (getIR i))) 
     	    irSensors)
     (printf "\n")
     (set! previousTime (current-inexact-milliseconds))
     )
  ) ;; end of print IR

  (cond ( (not (equal? (leftBump?) previousLeft))
          ;; Something has changed for the left bump
          ;; Just two cases: either it has been pressed, or released
          (cond ((leftBump?) (printf "Left bump pressed\n"))
                (else (printf "Left bump released\n"))
                )
          )
        ) ;; end of cond for left bump changed
  
  (cond ( (not (equal? (rightBump?) previousRight))
          ;; Something has changed for the right bump
          (cond ((rightBump?) (printf "Right bump pressed\n"))
                (else (printf "Right bump released\n"))
                )          
          )
        ) ;; end of cond for right bump changed
  
  ;; Set the state before iterating
  (set! previousLeft (leftBump?))
  (set! previousRight (rightBump?))
  
  (sleep 0.02)

  ;; A little trick to exit when both bump sensors are pressed
  (cond ((not (and (leftBump?) (rightBump?)))
         (controlLoop) 
         )
        )
  )

(define (minimalLoop)
  (open-asip)
  
  ;; let's take things easy...
  (sleep 0.2)
  (enableIR 100)
  (sleep 0.2)
  (enableBumpers 100)

  ;; half a second to stabilise
  (sleep 0.5)
  
  (controlLoop)
  (close-asip)
  )

\end{lstlisting}
The students also become familiar with the trial and error aspects of programming with real-time systems, such as the need for the sleep commands in the above code.

Other higher-order functions can also be employed by students in robotic applications. For example, an application might log a list of the moves a robot makes in exploring an environment under some algorithm such as that in Section \ref{section:random}. Filter functions might then be used with predicate arguments to extract interesting data, such as the number of right turns or the number of straight paths taken for more than a given time before hitting a wall. Fold functions might be used to process the data in a number of ways. The following examples show how students might use \texttt{map}, \texttt{filter} and \texttt{foldr} in working with the robots. 

Firstly, let us suppose that we want to read some Arduino input pins and find our how many  of them are set to high. The following code fragments illustrate this.

\begin{lstlisting}
;defined in AsipMain.rkt
(define HIGH 1)
(define LOW 0)

(define INPUTPINS (list 2 3 4))

;replace pin numbers with pin values
(map (lambda (i) (digital-read i)) INPUTPINS)

;count HIGH values
(length (filter (lambda (i) (= i HIGH)) INPUTPINS))

;alternative count using foldr
(foldr + 0 (filter (lambda (i) (equal? i HIGH)) INPUTPINS))
\end{lstlisting}

As a further example of the use of \texttt{foldr}, we return to the code that printed the values of the IR sensors at 3 second intervals and modify it so that instead of printing the IR values, they are accumulated in a list:

\begin{lstlisting}
;list of IR values, initially empty
(define IRlog (list))

;; snippet modified to Log IR values
  (cond ( (> (- currentTime previousTime) interval)
          ;; We use map to add IR values to a list
          (map (lambda (i) (cons (getIR i) IRlog)) irSensors)
          (set! previousTime (current-inexact-milliseconds))
          )
        ) 
\end{lstlisting}

The list could then be processed with \texttt{foldr} to find out things such as the sum of those IR readings greater than some threshold value.

\begin{lstlisting}
(define sumIRgreaterthan (lambda (threshold) 
        (foldr 
            (lambda (x y) (cond 
               ((> x threshold) (+ x y))
               (#t y))) 
             0  IRlog)))
\end{lstlisting}
%3.3 filter (must find example)
%3.4 foldl  foldr?
%3.5 Other patterns? currying in Racket is a pain...

\subsection{Contracts}\label{Section:contracts}

A contract in Racket is a promise that a developer makes about a piece of code. Racket contracts are typically defined for modules \cite{racket_mod}, collections of definitions that are then used by other Racket programs using the construct \texttt{(require modulename.rkt)}. The \texttt{(provide [...])} block is used to specify the definitions that are accessible when the module is included with a \texttt{(require )} statement. The role of contracts is explored by students first in a non-physical context (the creation of a bank account module), and then in the cyber-physical context of MIRTO to determine requirements on the robot's behaviour, for example its speed with respect to the hardware specification:

\begin{lstlisting}
(provide (contract-out ;; Begin of contract
          [speed (and/c number? exact-nonnegative-integer?)]
          [added_speed (-> checkSpeed any)]
          [current_speed (-> number?)]
          ) ;; End of contract
         )
 
;; We start from an initial speed of 0
(define speed 0)

;; added speed takes a value and adds it to the initial speed
(define (added_speed value) (set! speed (+ value speed)))

;; current speed returns the new value of speed
(define (current_speed) total)

(define checkSpeed
  (lambda (a) 
    (and (number? a) (integer? a) (exact? a) 
    (and (>= (+ a total) -255) (<= (+ a total) +255))
    )
  )
\end{lstlisting}

We shall justify briefly in Section \ref{Section:lit} the value of this specific construct for the purposes of learning programming.

\section{Robotic examples by students}\label{Section:examples}

The final project work for the year consisted of the design of interesting applications for the robots, which some students tackled with much skill and imagination. One team had their robot `race' against falling dominos, following identical paths (\url{https://www.youtube.com/watch?v=RnzDDdN0B14}). Another team implemented a PID algorithm that used the values of the robot's three IR sensors to follow lines drawn on a surface (\url{https://www.youtube.com/watch?v=VKXLM4av54o}); another team created two robots of their own and entered them in the Eurobot national championships in April 2015, coming 4th out of 17 teams (\url{https://youtu.be/o8b63XqIg5Y}). Such achievements were unheard of in the predecessor of the current CS programme, and much of this success is the result of the motivation instilled in the students by the opportunity to apply their developing knowledge and skills to real-world problems through using the robots.

In the line-following project, students started studying the design principles of open- versus closed-loop systems, to understand how to feedback values from sensors in the code for other actuators. This was followed by the study of mathematical principles to design first a ``bang-bang'' line-following algorithm, then improved to a proportional controller to change the speed of the wheels, finally extended to a proportional-integral-derivative controller. At least one team of students did extended testing, both of the code and of its execution on MIRTO to find the optimal setting for various tracks, see \url{https://www.youtube.com/watch?v=VKXLM4av54o}. Here below we present their code, construed around the various functions of the IR-sensors and PID-controller that we helped them define. This code was developed autonomously by the team, without any external help from the tutors.

\begin{lstlisting}
(define previousTime ( current-inexact-milliseconds ))
(define currentTime 0)

(define interval 10)

;; The list of IR sensors (used in map below)
(define irSensors (list 0 ))
(define irSensors1 (list 1))
(define irSensors2 (list 2))

(define (irLoop)
  (set! currentTime ( current-inexact-milliseconds ))
  (cond ( (> (- currentTime previousTime ) interval )
          (set! previousTime ( current-inexact-milliseconds ))
          )
        )
  (irLoop))

(define (IRsweg a b c)
  (define curRightCount (getCount 0))
  (define curLeftCount (getCount 1))
  (define (searchLoop)
    (set! curRightCount (getCount 0))
    (set! curLeftCount (getCount 1))
    (cond ( (or (< 45 (getIR a)) (< 45 (getIR b)) (< 45 (getIR c)))
		        (stopMotors))
          ( (or (>= curRightCount 16) (<= curLeftCount -16)) (stopMotors)
            (sleep 0.1)
            (setMotors -115 -115)
            (sleep 0.1)
            (cond ((or (< 45 (getIR a)) (< 45 (getIR b)) (< 45 (getIR c)))
						  (stopMotors))
            (#t (searchLoop) )
                             )
                             )
          (#t (printf "~a ~a\n" (getCount 0) (getCount 1)) (searchLoop))
          )                                                        
    )
  
(define (search)
  (resetCount 0)
  (resetCount 1)
  (setMotors 115 115)
  (sleep 0.1)
  (searchLoop)
    )
  
 (cond
    ((and (> 45 (getIR a)) (> 45 (getIR b)) (> 45 (getIR c))) (search))
    ((and (< 45 (getIR a)) (< 45 (getIR b)) (< 45 (getIR c)))
		  (setMotors -115 115))
    ((and (< 45 (getIR a)) (< 45 (getIR b)) (> 45 (getIR c)))
		  (setMotors 0 115))
    ((and (> 45 (getIR a)) (< 45 (getIR b)) (< 45(getIR c)))
		  (setMotors -115 0))
    ((and (< 45 (getIR a)) (> 45 (getIR b)) (> 45 (getIR c)))
		  (setMotors 0 115))
    ((and (> 45 (getIR a)) (> 45 (getIR b)) (< 45 (getIR c)))
		  (setMotors -115 0))
    (#t (setMotors 0 0)))
  (IRsweg a b c)
  )

(define cIR 
  (lambda (i) 
     (cond ( (> (getIR i) 45)
            (getIR i)
            )
          (else 0)
          )
    )
  )

(define oldError 0)
(define speed 150)
(define sumError 0)
(define (IRsweggier a b c)
  (define Kp 0.05)
  (define Kd 0.045)
  (define Ki 0.007)
  (define currentError 0)
  (cond [ (> (+ (cIR a) (cIR b) (cIR c)) 0)
           (set! currentError (/ (+ (mult 0 (cIR a)) (mult 2000 (cIR b))
					   (mult 4000 (cIR c))) (+ (cIR a) (cIR b) (cIR c))))]
           [else 
           (cond [ (> oldError 2000)
	    (set! currentError 4000)]
               (else (set! currentError 0))
               )
         	]
        )
  (define correction (inexact->exact (round (+ (mult Kp (- currentError 2800)) 
                                     (mult Kd (- currentError oldError) )
                                     (mult Ki sumError)
                                     )))
    )
  (displayln currentError)
  (displayln correction)
  
  (cond
     ((< correction 0) (setMotors (- (+ speed correction)) speed) 
     )
    ((> correction 0) (setMotors (- speed) (- speed correction))
    )
    (\#t (setMotors (- speed) speed))
    )
  (sleep 0.02)
  (set! oldError currentError)
  (cond 
        ( (and (> currentError -400) (< currentError 400)) (set! sumError 0))
        (else (set! sumError (+ sumError (- currentError 2000))))
  )
  (IRsweggier a b c)
  )
\end{lstlisting}

%  Line following
%  Videos and pictures
%  
%  We had a competition towards the end of the course for teams to design interesting applications for the robots, which some students tackled with much skill and imagination. One team had their robot `race' against falling dominos, following identical paths. Another team implemented a PID algorithm that used the values of the robot's three IR sensors to follow lines drawn on a surface. The best team, however, created two robots of their own and entered them in the Eurobot national championships in April 2015, coming 4th out of 17 teams. Such achievements were unheard of in the predecessor of the current CS programme, and much of this success is the result of the motivation instilled in the students by the opportunity to apply their developing knowledge and skills to real-world problems through using the robots.
%  
%  VIDEOS AVAILABLE AT???
%  

\section{Related literature}\label{Section:lit}

The principles at the basis of our First Year BSc in Computer Science highlighted in Section \ref{Section:intro} reflect much of the current literature in pedagogy, where we broadly follow a fine-grained, outcome-based learning path model. The theoretical implications remain to be assessed in their full meaning, especially for the pedagogical support; see \cite{DBLP:journals/tsmc/YangLL14} for a recent overview. However, this approach also follows professional guidelines and advice from industry. For example, the ACM/IEEE 2013 CS 2013 Curricula \cite[p.28]{CS2013} in section 4.1 discourages 

\begin{quote}
``to associate each Knowledge Area with a course [\dots] even though many curricula will have some courses containing material from only one Knowledge Area or, conversely, all the material from one Knowledge Area in one course. We view the hierarchical structure of the Body of Knowledge as a useful way to group related information, not as a structure for organizing material into courses. Beyond this general flexibility, in several places we expect many curricula to integrate material from multiple Knowledge Areas''.
\end{quote}
The structure of our course reflects precisely this principle and, albeit we could have used a purely imperative approach to obtain the same final results in terms of robotic applications, we have chosen to emphasise functional programming, a topic represented as one of the first three Knowledge Units in the ACM/IEEE CS Curricula. In our approach, we try to cover all the Core-Tier1 and Core-Tier2 Topics from the Curriculum, while our assessment methodology (see \cite{ELS}) focuses explicitly on all of the Core-Tier1 and Core-Tier2 Learning Outcomes.

However, the essential role of functional programming in teaching is highlighted not only in the academic context. In a very recent contribution \cite{ballzornACM}, it is stressed how programmers should be exposed as early as possible to functional programming also as a way to gain exposure in declarative language abstractions, and how this principle is highly appreciated in the industry. This nicely complements the richness of functional constructs (with their imperative flavour mentioned at the beginning of Section \ref{Section:RacketMirto}) that Racket allows us to teach to our students.

Another important recommendation from \cite{ballzornACM} for teaching programming concerns ``design by contracts'' as a way to refer to annotations made in the program to express what the program (or part of it) is supposed to accomplish, as opposed to how it should compute. This technique falls in the larger and more essential issue of educating the future generation of computer scientists and programmers with 

\begin{quote}
``a grounding in logic, its application in design formalisms, and experience the creation and debugging of formal specifications'' \cite[p.31]{ballzornACM}.
\end{quote}
Besides our coverage of formal topics (including functions, relations, set theory, regular expressions, propositional and predicate logic) and our design workshops (in which essential topics such as UML and (extended) finite state machines are introduced), we implement directly the design by contracts principle in Racket, as illustrated in section \ref{Section:contracts}.

There are other works dealing with the subject of robots and functional
programming. The approach used in~\cite{carlsonrobotlang} and~\cite{Wakeling2008} is quite similar to ours, though the former uses functional programming
to teach robot operation, rather than the other way around. The approach
in~\cite{hudak2003} is more advanced and uses functional reactive programming.
This is a concept that, while interesting, we feel is not a topic for a basic
first year programming course.

\section{Conclusion}

We have introduced an overview of how the use of Racket to drive the MIRTO robot lends itself to teaching functional patterns to first year students, and the role of these in our first year Computer Science curriculum. 

Our chief goals in using robots in the curriculum were fivefold.
\begin{itemize}
	\item To teach some real-time robotics programming.
	\item To reinforce the learning of functional and imperative programming
	\item To help students to develop their practical group-working and project skills
	\item To enable students to reinforce and integrate the knowledge and skills acquired in other parts of the curriculum
	\item To encourage the students to explore beyond the confines of their programme of study, by competing with other students within the university and from other universities.
\end{itemize}

The programming ranged from the application of functional programming concepts, such as higher-order functions, lists of functions and vectors, to the use of low-level imperative programming such as insertion of delays. The latter could have been hidden behind abstractions but exposing the students to such concepts directly gave them a broader appreciation of the many facets of practical problem solving in programming.

In essence, we have used functional programming as a tool not only to reinforce the teaching of most aspects of the curriculum, but also to introduce students to robotics applications. On the other hand, we have also used robotics to deepen students' knowledge and skills in functional programming. It is true that, in terms of final results for the implementation of robotic applications, all our functional patterns could have been converted to imperative ones.
However, students are excited by working with the robots, and this excitement tends to make them engage more and thereby to achieve more in a topic (functional programming) that is normally considered ``theoretical''. The ability to motivate is undoubtedly one of the most important aspects of the robotics projects, especially when compared with the rather dry applications normally included in first-year and functional programming courses. At the time of writing, two cohorts of students have passed through the first year of the programme, and pass rates for those students completing the year were over 90\% in each case. We believe this success to be due in no small part to the motivating influence of the practical projects undertaken by the students.

\nocite{*}
\bibliographystyle{eptcs}
\bibliography{generic}
\end{document}